\documentclass[sigplan,screen]{acmart}

\AtBeginDocument{%
  }

\setcopyright{none}
\renewcommand\footnotetextcopyrightpermission[1]{}

\copyrightyear{2026}
\acmYear{2026}
\acmDOI{XXXXXXX.XXXXXXX}
\acmConference[InSyDe workshop @ EuroSys '26]{Make sure to enter the correct
  conference title from your rights confirmation email}{April 27,
  2026}{Edinburgh, UK}

\begin{document}

\title[LLM-Driven Design Space Exploration of FPGA-based Accelerators]{LLM-Driven Design Space Exploration of \\ FPGA-based Accelerators}

\author{Vinamra Sharma, Xingjian Fu, Jude Haris, Jos\'e Cano \\
\emph{School of Computing Science, University of Glasgow, Scotland, UK}
}

\renewcommand{\shortauthors}{Sharma et al.}


\begin{abstract}

Designing field-programmable gate array (FPGA)-based accelerators for modern artificial intelligence workloads requires navigating a large and complex hardware design space encompassing architectural parameters, dataflow strategies, and memory hierarchies, making the process time-consuming and resource-intensive. 
While the SECDA methodology enables rapid hardware–software co-design of accelerators through SystemC simulation and FPGA execution, identifying optimal accelerator configurations still requires substantial manual effort and domain expertise. 
This work presents SECDA-DSE, a framework that integrates Large Language Models (LLMs) into the SECDA ecosystem, comprising tools built around SECDA to automate the design space exploration (DSE) of FPGA-based accelerators. 
SECDA-DSE combines a structured DSE Explorer for generating accelerator configurations with an LLM Stack that performs reasoning-guided exploration using retrieval-augmented generation and chain-of-thought prompting, alongside a feedback loop that enables reinforced fine-tuning for continuous improvement. 
We demonstrate the feasibility of SECDA-DSE through an initial high-level synthesis based evaluation of a generated accelerator design that meets synthesis timing and resource constraints on an Zynq-7000 FPGA.

\end{abstract}




\setcounter{topnumber}{5}
\setcounter{totalnumber}{10}
\renewcommand{\topfraction}{0.95}
\renewcommand{\textfraction}{0.05}

\setlength{\textfloatsep}{6pt}
\setlength{\floatsep}{6pt}
\setlength{\intextsep}{6pt}

\maketitle

\section{Introduction}
\label{intro}

Modern artificial intelligence (AI) workloads rely on specialized hardware accelerators to achieve high performance and energy efficiency. 
field-programmable gate arrays (FPGAs) have emerged as an attractive option for this purpose due to their reconfigurability and ability to provide customized computation pipelines tailored to specific AI workloads. 
However, designing efficient FPGA-based accelerators for AI workloads remains a challenging and time-consuming process that requires extensive expertise in hardware-software co-design, high-level synthesis (HLS), and system integration.

AI accelerators comprise architectural parameters, data-flow strategies, and memory hierarchies; together, these factors create a large hardware design space that must be explored during the design process.
This exploration is often performed manually or supported by a limited number of automated tools~\cite{cesaretti2025rapid, tiwari2025hardware}, leading to prolonged development cycles and effort.
Furthermore, each design iteration may require costly HLS and evaluation stages, slowing down the exploration of the hardware-software co-design space~\cite{gibsonDLASConceptualModel2025}.

The SECDA (SystemC Enabled Co-Design of DNN Accelerators) methodology~\cite{haris2021secda} was proposed to reduce the development time and effort of FPGA-based accelerators by enabling hardware-software co-design through SystemC simulation and hardware synthesis workflows. 
This approach allows developers to rapidly iterate on accelerator designs while maintaining a unified development flow.

Building upon the SECDA TFLite, the SECDA-DSE framework aims to further automate the accelerator design process by incorporating design space exploration (DSE) techniques through the effective use of large language models (LLMs).
SECDA-DSE leverages the reasoning capabilities of LLMs to guide the exploration of hardware design parameters, generate candidate accelerator architectures, and iteratively refine designs based on evaluation and performance feedback. 
We discuss the framework architecture, including the two major component, the DSE Explorer and the LLM Stack, and outline how these components enable iterative exploration, evaluation, and refinement of new designs. 
We present the vision and early design of SECDA-DSE along with preliminary results through verification of a generated accelerator design through HLS on a Zynq-7000 FPGA.


\section{Background and Related Work}
\label{background}

SECDA (SystemC Enabled Co-Design of DNN Accelerators) is a hardware-software co-design methodology aimed at reducing the development time of FPGA-based accelerators for DNN inference~\cite{haris2021secda}. It provides accelerator design, system simulation, and application framework integration within a unified development environment. 

SECDA-TFLite~\cite{haris2023secda} extends the SECDA methodology by providing a toolkit for integrating FPGA accelerators into the TensorFlow Lite (TFLite) framework directly. It includes modules for SystemC simulation, profiling, and AXI-based communication, enabling developers to rapidly prototype and evaluate new accelerator designs. Through the use of the TFLite delegate system, SECDA-TFLite allows portions of a DNN model to be offloaded to custom hardware accelerators while maintaining compatibility with the existing inference pipeline. 

Despite these advances, the process of identifying efficient accelerator designs remains challenging and time consuming even for experts, since it involves exploring numerous architectural parameters such as compute unit dimensions, memory organization, tiling strategies, and dataflow patterns. Manually exploring these parameters needs multiple iterations and often relies on domain expertise~\cite{dhilleswararao2022efficient, boutros2025field}. 
However, the recent progress in Large Language Models (LLMs) offers new opportunities to automate parts of the DSE process without the need for extensive domain expertise~\cite{ling2025domain, zheng2025automation}. 
For example, iDSE~\cite{li2025idse} proposes an LLM-based framework for navigating HLS design spaces, where the model guides the search process and prunes candidate configurations to converge more efficiently toward optimal solutions. 
Similarly, LUMINA~\cite{zhang2026luminallmguidedgpuarchitecture} proposed an LLM-driven architecture exploration framework for GPU design, where the model analyzes simulator code and performs bottleneck-based reasoning. 
Also, LIMCA~\cite{vungarala2025limca} proposes an LLM-driven framework to explore analog in-memory computing architectures, where the model generates and evaluates configurations. 

While these works demonstrate the growing potential of LLMs in assisting hardware architecture exploration across domains, SECDA-DSE differs by (i) integrating LLM reasoning directly within the accelerator design workflow and (ii) enabling iterative DSE guided by a hardware evaluation feedback loop.

\section{SECDA-DSE}
\label{method}

\begin{figure}[!tb]
  \centering
  \includegraphics[width=0.89\linewidth]{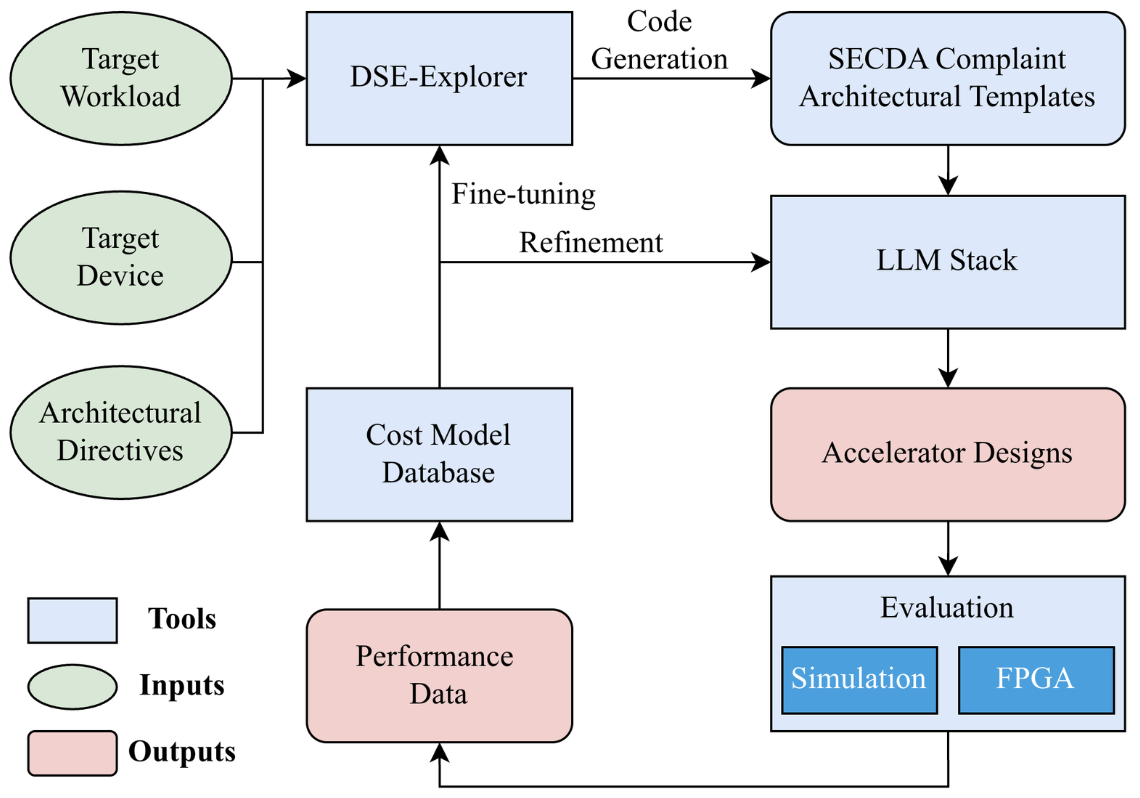}
  \caption{SECDA-DSE framework architecture showing the interaction between the DSE Explorer and the LLM Stack.}
  \label{fig:SECDA-DSE Overview}
  \Description{SECDA-DSE framework architecture showing the interaction between the DSE Explorer and the LLM Stack.}
\end{figure}

SECDA-DSE extends the SECDA ecosystem, which comprise of various tools and components present in SECDA including AXI APIs, hardware monitors and simulation profiler by introducing an automated framework for exploring hardware accelerator design spaces.
Figure~\ref{fig:SECDA-DSE Overview} shows the overall architecture of SECDA-DSE, which consists of two main components: (i) \textbf{DSE Explorer}\textbf{:} Responsible for generating and evaluating hardware configurations; (ii)  \textbf{LLM Stack}\textbf{:} Responsible for reasoning over the design space and guiding exploration strategies.

The framework takes as inputs: A target AI workload (e.g., CNNs, LLMs); A target FPGA device (e.g., PYNQ-Z1, Kria KV260); Architectural directives for exploration (e.g., tiling strategies, compute unit dimensions, memory hierarchy parameters). 
Based on these inputs, SECDA-DSE generates candidate accelerator architectures and evaluates them using the SECDA ecosystem.
The evaluation results are then further used to refine subsequent design iterations through a human-in-the-loop feedback for reinforcement.

More specifically, the workflow begins with user inputs that are provided to the DSE Explorer, which generates hardware configurations and produces SECDA-native architecture that follows SECDA defined architectural guidelines and patterns.
The templates are then processed by the LLM Stack, which performs reasoning-driven refinement of the design space and generates the final accelerator designs.

The resulting designs are evaluated through simulation using the SECDA-TFLite toolkit~\cite{haris2023secda}, producing performance metrics such as execution latency, resource utilization, and data movement costs. These metrics are collected as performance data and stored in a cost model database, which is then used to guide subsequent exploration iterations through refinement and fine-tuning. By incorporating this feedback-driven loop, SECDA-DSE enables iterative improvement of accelerator configurations.

To support this iterative exploration process, SECDA-DSE is designed as a modular orchestration framework in which each component exposes an API endpoint for data interchange. The DSE Explorer generates candidate architectural parameter sets (e.g., tiling factors, compute array dimensions, memory allocation parameters) and instantiates them into SECDA-compliant accelerator templates. These templates are then passed to the LLM Stack, which retrieves relevant SECDA-TFLite implementation context, reasons over prior hardware datapoints, and proposes refined candidate configurations. Each generated design is evaluated through SECDA-based simulation and, where required, downstream synthesis and hardware execution flows, with the resulting performance and resource metrics stored in the cost model database for reuse in subsequent iterations.

The DSE Explorer and the LLM Stack interact iteratively rather than sequentially. The DSE Explorer proposes candidate parameter permutations and produces summarized hardware data-points, while the LLM Stack consumes these data-points together with retrieved implementation context to rank, refine, or reject candidate designs before the next exploration round. This interaction enables SECDA-DSE to combine structured parameter exploration with reasoning-guided design refinement while also harnessing a Chain-of-Thought (CoT) prompting approach which enforces the LLM to take instructions and step by step.


\subsection{DSE Explorer}

\begin{figure}[!tb]
  \centering
  \includegraphics[width=0.8\linewidth]{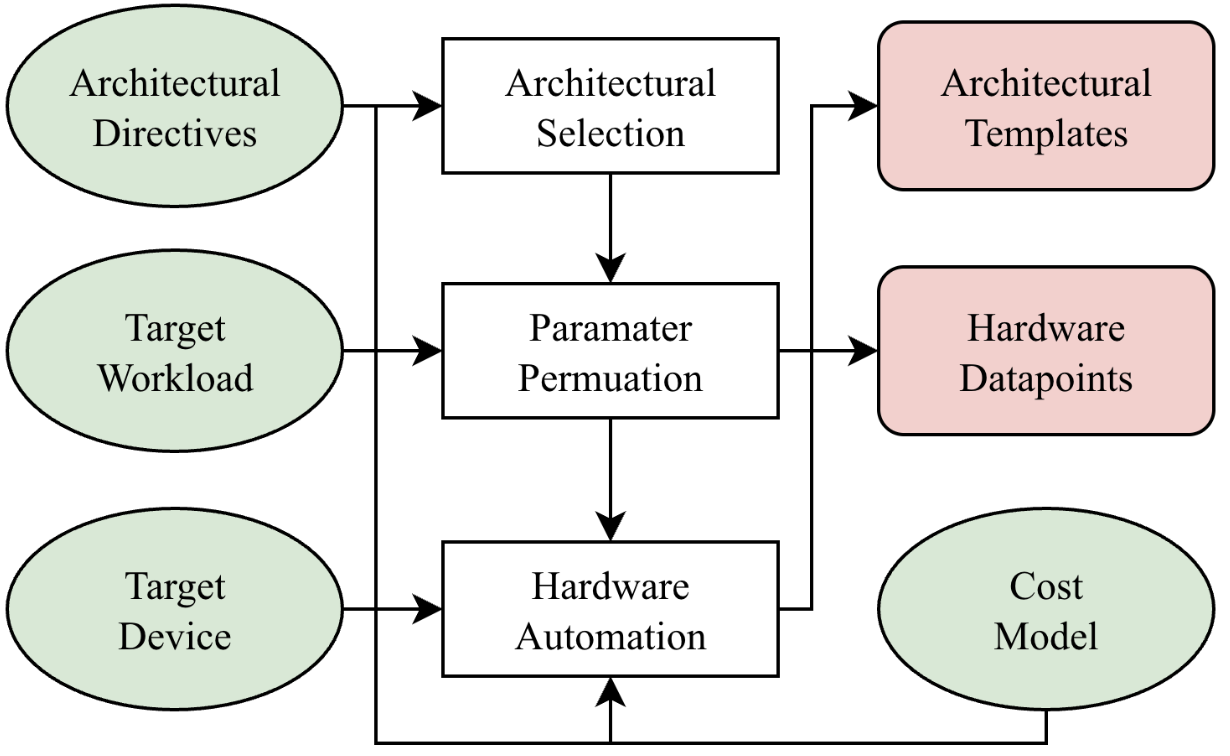}
  \caption{Workflow of the DSE Explorer.}
  \label{fig:DSE-Explorer}
  \Description{Workflow of the DSE Explorer.}
\end{figure}

The DSE Explorer performs the automated exploration of the accelerator design space.
It is responsible for generating and evaluating candidate accelerator configurations within the SECDA-DSE framework. As shown in Figure~\ref{fig:DSE-Explorer}, its primary objective is to navigate the large hardware design space associated with FPGA-based AI accelerators by combining structured exploration with feedback from evaluation results.

The DSE explorer operates based on three primary inputs: the target AI workload, the target FPGA device, and a set of architectural directives that constrain or guide the exploration process.
Using this information, it generates accelerator configurations that conform to SECDA-native architectural templates, enabling integration with the existing SECDA design ecosystem. 
The DSE Explorer works by taking in an accelerator design (generated by the LLM Stack or initially by an expert designer) with pre-defined hardware parameters and exploring the permutations suggested by the LLM Stack.
Each permutation generates a design run folder that includes the source code, HLS-generated RTL code, and the FPGA-mapped design.
The FPGA-mapped design is the evaluated design, and performance metrics are collected. 
Alongside this, the resource utilization of the accelerator design is used to generate a summarized set of "results" for each of the permutation run.
The summarized results along with the source code for the run are later fed into the LLM Stack as part of the "hardware data points".

\subsection{LLM Stack}

The LLM Stack acts as an intelligent reasoning and orchestration layer within SECDA-DSE, guiding the exploration of the hardware design space generated by the DSE Explorer. Its primary objective is to analyze accelerator configurations, reason about architectural trade-offs, and refine exploration strategies based on evaluation feedback and historical performance data logs.

\begin{figure}[!tb]
  \centering
  \includegraphics[width=0.99\linewidth]{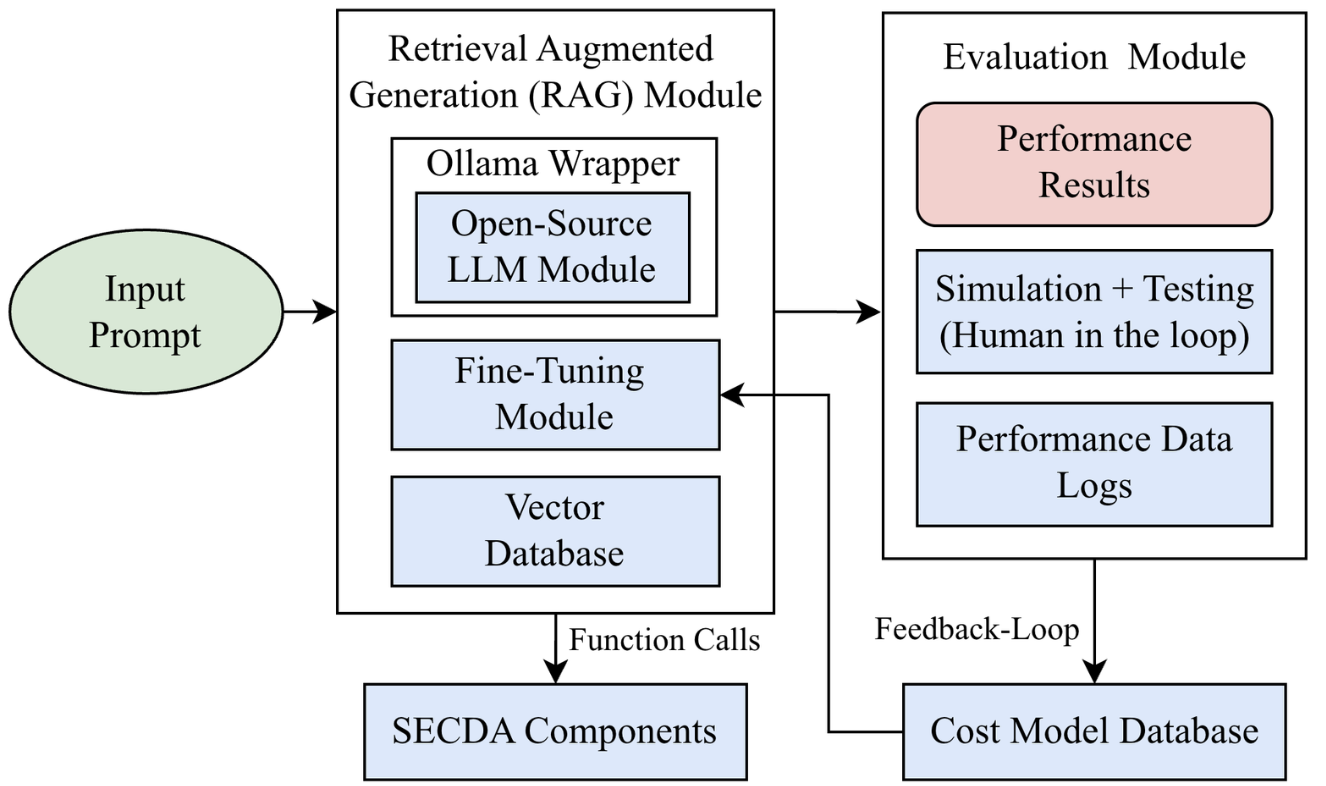}
  \caption{LLM Stack Architecture illustrating various components and interactions.}
  \label{fig:LLM STACK}
  \Description{LLM Stack Architecture}
\end{figure}

Figure~\ref{fig:LLM STACK} provides a block representation of its various modules and their interactions within and across the stack. The LLM Stack consists of two major modules: the Retrieval Augmented Generation Module (RAG) and the Evaluation module with external Model Context Protocol (MCP)-based API calls to access SECDA components.

\subsubsection{Retrieval Augmented Generation Module (RAG)}

The full LLM Stack pipeline is developed in a modular way using the open-source framework Ollama~\cite{ollama} for inference on-device. Ollama also enables switching between newer LLMs with ease without requiring any additional code changes for integration, making the tool easy to use for the end-user.

RAG allows the LLM to retrieve relevant information and have enough context of SECDA through a vectorized database consisting of the SECDA-TFLite code-base~\cite{haris2023secda} indexed for search through added comments. 
The RAG module does not expose the full SECDA-TFLite codebase or complete raw hardware logs at each iteration. Instead, it retrieves only the most relevant code fragments, template definitions, and API-level context required for the current design decision. This approach allows to maintain token limit while providing enough context to the LLM for making informed decisions.

Note that the LLM Stack also incorporates a parameter-efficient fine-tuning module, which adapts the LLM using data generated during the exploration process. Performance metrics obtained from simulation and FPGA execution are collected as the hardware data points and stored within the cost model database. 
In this setting, the fine-tuning dataset is constructed from previously explored accelerator designs and their associated evaluation outcomes. Each training data point includes the proposed architectural configuration, workload and device context, and the resulting feedback signals including simulation success, latency, and resource utilization. These signals enable supervision for adapting the model toward generating configurations that are both performance centric and feasible.

\begin{figure}[!tb]
  \centering
  \includegraphics[width=0.98\linewidth]{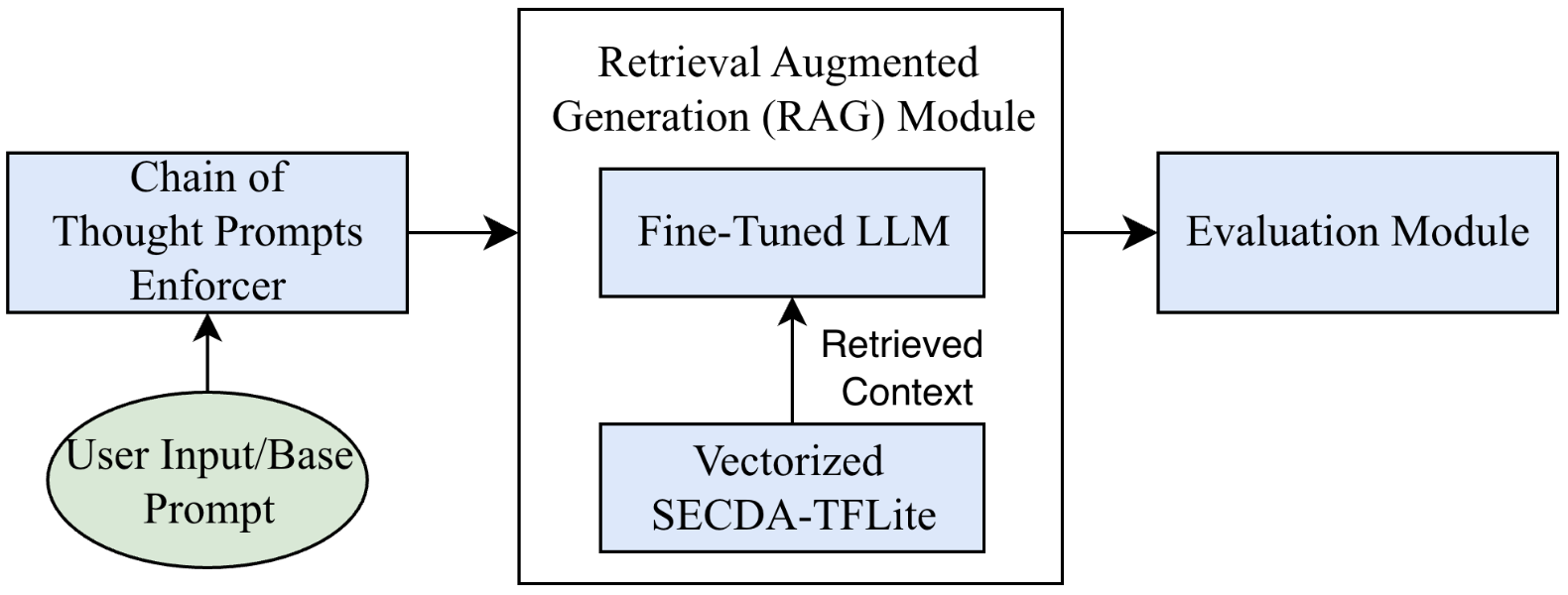}
  \caption{Block diagram showing how Chain-of-Thought (CoT) component is used for improving prompt structure.}
  \label{fig:COT}
  \Description{Block diagram showing how Chain-of-Thought (CoT) component is used for improving prompt structure}
\end{figure}

In addition to retrieval-based grounding, the LLM Stack employs CoT prompting as shown in Figure~\ref{fig:COT}. CoT enables structured and multi-step reasoning when exploring architectures and guides the LLM to generate intermediate reasoning steps before producing final outputs. This aims to improve its ability to analyze complex design constraints and architectural trade-offs without the extensive need for a bigger model with a pre-enabled thinking node available.

\subsubsection{Evaluation module}

The Evaluation module is responsible for assessing the performance and feasibility of the generated accelerator designs. Its primary role is to provide feedback that guides the iterative DSE process, as shown in Figure~\ref{fig:LLM STACK}. The module evaluates each accelerator configuration through a combination of simulation-based analysis and hardware execution initially having a human-in-the-loop but aimed to further automate and take the human-out-of-the-loop once the data-log size grows, enabling quick validation of design decisions.

The evaluation workflow begins with SystemC-based simulation, which allows the accelerator architecture to be tested within a high-level simulation environment. SystemC simulation enables rapid verification of functional correctness and provides early estimates of performance metrics such as execution latency, hardware resource estimation, and the data movement costs. These results from executions are then saved in the database, which are further passed for fine-tuning the base LLM.

To reduce invalid design proposals, SECDA-DSE constrains design generation through SECDA-compliant architectural templates and device-aware parameter ranges rather than allowing unconstrained free-form design generation. 
Candidate accelerator configurations are therefore bounded by workload requirements, and architectural directives before evaluation. Further, as shown in ~\ref{fig:LLM STACK}, there is a human-in-the-loop for testing the generated design. Any infeasible or invalid configurations that fail simulation, violate hardware resource limits, or cannot progress through downstream tool flows are rejected and logged as negative hardware data points for future refinement.

Note that the LLM Stack uses Low-Rank Adaptation (LoRA) as a parameter-efficient fine-tuning mechanism. LoRA~\cite{hu2022lora} is a widely used technique for adapting LLMs without updating the full set of model parameters. 
Instead of retraining the entire model, LoRA freezes the original pretrained weights and introduces small trainable low-rank matrices into the selected layers of the LLM architecture, enabling task-specific adaptation with fewer trainable parameters.


To reduce the risk of premature convergence to locally optimal designs, SECDA-DSE maintains exploration diversity across iterations instead of focusing only on the current best-performing configuration. It does so by evaluating multiple candidate parameter permutations, refining the search based on both successful and unsuccessful hardware data points, and enabling the LLM Stack to reason over a broader range of design trade-offs before generating new accelerator designs for the target AI workload.

\section{Preliminary Results}
\label{Early-Results}

We conducted an initial evaluation to validate the first step of the SECDA-DSE loop: translating a natural language accelerator specification (see Appendix) into a SECDA-native implementation that can be compiled through HLS. 
The LLM Stack was provided with a prompt describing an element-wise vector multiplication kernel, where two input vectors $X$ and $Y$ of length $L$ are streamed via AXI-Stream interfaces into on-chip buffers. The accelerator follows a load-compute-store model, where the compute stage performs $Z_i = X_i \odot Y_i$ and streams the output vector $Z$ back to memory. 

Overal, SECDA-DSE successfully generated a complete SECDA-native accelerator workspace, including the SystemC accelerator description, build integration, and a software driver.


\subsection{High Level Synthesis}

The evaluation of the generated hardware accelerator was performed using Vivado HLS 2019.2~\cite{vivadohls2019_2} targeting Zynq-7000 FPGA (\texttt{xc7z020-clg400 -1} SoC) with a clock constraint of $5.00$\,ns ($200$\,MHz) and an estimated critical path delay of $3.950$\,ns, meeting the specified clock constraint.


\subsection{Latency and Throughput}

The overall accelerator latency ranges from $0$ to $2060$ cycles, corresponding to $0$\,ns to $10.300\,\mu$s at $200$\,MHz. The initiation interval (II) also ranges from $0$ to $2060$ cycles, indicating that the design is not fully pipelined at the top level. 
Table~\ref{tab:vecmul-latency} summarizes the latency characteristics of the submodules.

\begin{table}[t]
  \centering
  \caption{HLS latency and initiation interval (II).}
  \label{tab:vecmul-latency}
  \begin{tabular}{lccc}
    \hline
    Module & Latency (cycles) & II (cycles) & Pipeline \\
    \hline
    HW\_MAIN & 3-3 & 3-3 & None \\
    Send     & 7-1030 & 7-1030 & None \\
    Compute  & 13-1036 & 13-1036 & None \\
    Recv     & 8-2059 & 8-2059 & None \\
    \hline
  \end{tabular}
\end{table}


\begin{table}[t]
  \centering
  \caption{HLS resource utilization.}
  \label{tab:vecmul-resources}
  \begin{tabular}{lccc}
    \hline
    Resource & Used & Available & Utilization (\%) \\
    \hline
    BRAM\_18K & 6 & 280 & $\sim$2 \\
    DSP48E    & 3 & 220 & $\sim$1 \\
    FF        & 993 & 106{,}400 & $\sim$0 \\
    LUT       & 1113 & 53{,}200 & $\sim$2 \\
    \hline
  \end{tabular}
\end{table}

\subsection{Resource Utilization}

Table~\ref{tab:vecmul-resources} shows the estimated utilization of the FPGA resources reported by HLS. The compute module accounts for the majority of arithmetic resources, utilizing 3 DSP units along with moderate LUT and FF usage. 
The design maps three on-chip buffers (for $X$, $Y$, and $Z$), each implemented using BRAM. 
These preliminary results demonstrate that SECDA-DSE is capable of generating SECDA-compliant accelerator designs from natural language specifications into SECDA-compliant designs that successfully traverse the HLS toolchain. 
Note that the results are based on HLS estimates, and further evaluation and FPGA execution are part of the full SECDA-DSE scope discussed in Section~\ref{Discussions}.

\section{Next Steps}
\label{Discussions}

This section identifies three key next steps for advancing SECDA-DSE: (i) full system integration through MCP-based automation; (ii) comprehensive evaluation of both the end-to-end system and its individual components; and (iii) large-scale validation and benchmarking across multiple AI workloads, hardware platforms, and LLM configurations. 
These directions are pursued while accounting for key challenges, including the computational cost of large-scale exploration, the dependence on high-quality hardware data for effective learning, and the risk of suboptimal design generation. Next, we discuss each of these directions and challenges in detail.

\subsection{Full system integration} 

We plan to integrate full SECDA-DSE automation by connecting the DSE Explorer component with the LLM Stack through an MCP library. This library will expose SECDA-based simulation, hardware automation, the collection of generated hardware designs, and the associated resource utilization metrics per design. 
Exposing simulation endpoints will allow the LLM Stack to iteratively generate and evaluate accelerator designs without requiring full hardware synthesis, enabling efficient exploration within the SECDA simulation-driven workflow. In this setting, the LLM Stack effectively replaces the human expert within the simulation-based design loop. 


\subsection{Comprehensive evaluation} 

To evaluate the effectiveness of SECDA-DSE, we aim to establish a comprehensive evaluation strategy that considers both end-to-end system performance and individual subsystem contributions. 
The DSE Explorer will be evaluated based on search efficiency and parameter space coverage, while the LLM Stack will be assessed in terms of the quality of proposed refinements, validity of generated configurations, and effectiveness of retrieval-based reasoning. 


\subsection{Large-scale validation and benchmarking} 

We plan to conduct large-scale validation and benchmarking of SECDA-DSE across diverse workloads, hardware platforms, and LLM configurations. This includes generating and evaluating accelerator designs for multiple AI workloads using hardware execution flows on multiple FPGA targets. 
We will explore multiple open-source LLMs (e.g., Llama, Qwen), both pre-trained and post fine-tuning, to assess performance gains or losses over time using extended datasets and cross-model validation. This comprehensive evaluation will enable systematic benchmarking of SECDA-DSE, support refinement of the fine-tuning methodology, and provide generalizability.



\subsection{Challenges} 

Despite its potential, SECDA-DSE also presents multiple challenges: (i) even simulation-based evaluation can remain computationally expensive when exploring large accelerator design spaces; 
(ii) the quality of the DSE process depends on the consistency and representativeness of the hardware data points collected used for retrieval and adaptation; 
(iii) although the SECDA framework constrains generation through templates and validation flows, LLM-guided reasoning may still produce suboptimal designs, particularly in underexplored regions of the design space. 
Addressing these challenges will be critical in future iterations of the framework, alongside expanding the hardware design database.

\section{Conclusion}
\label{conclude}

This paper introduced SECDA-DSE, a framework that extends the SECDA ecosystem with automated design space exploration for FPGA-based AI accelerators harnessing LLMs. 
SECDA-DSE combines two major components, the DSE Explorer and an intelligent LLM Stack, to guide the exploration of hardware accelerator architectures through reasoning and context driven decision making. 
SECDA-DSE enables the generation and refinement of hardware accelerator configurations based on workload requirements, device constraints, and architectural directives through retrieval-augmented generation. Furthermore, chain-of-thought reasoning, and parameter-efficient fine-tuning through LoRA enables efficient model adaptation by freezing the base model parameters and learning small low-rank matrices that specialize model to a new task, reducing computational and memory costs during fine-tuning. 
SECDA-DSE leverages LLMs to analyze hardware architecture trade-offs and iteratively improve generated accelerator designs through reinforcement guided by human-in-the-loop feedback. 
Our preliminary evaluation demonstrates that SECDA-DSE can successfully translate high-level natural language specifications into SECDA-native accelerator designs that meet HLS synthesis constraints, validating the feasibility of LLM-driven accelerator generation.

Future work will focus on fully automating the design space exploration loop by integrating the DSE Explorer and LLM Stack through an MCP interface, expanding the evaluation, and exploring the effectiveness of multiple fine-tuned open-source LLMs across larger hardware design datasets.


\begin{acks}


All authors gratefully acknowledge UK Research and Innovation (UKRI) and Engineering and Physical Sciences Research Council (EPSRC) funding for the AI Hub for Productive Research and Innovation in Electronics (APRIL) AI Hub [grant number EP/Y029763/1].
\end{acks}



\bibliographystyle{ACM-Reference-Format}
\bibliography{refs}


\appendix

\section*{Appendix}


The following is the initial prompt provided to the LLM stack for the generation of an element-wise vector multiplication accelerator design:

\emph{
I would like to create a hardware accelerator design.  
The accelerator should be able to take two input vectors: X and Y, both of length L.  
The accelerator should perform an element-wise multiplication operation and produce an output vector Z.  
The accelerator has two AXI-Stream based interfaces for loading X and Y data into custom X and Y buffers.  
The accelerator should also have a fixed length parameter L.  
Once the data is loaded, the accelerator should execute the element-wise multiplication in parallel and store the results in buffer Z within the compute module.  
The loading should be performed using a load module.  
Finally, the results should be written back to main memory using a store module that outputs via an AXI-Stream interface.  
Create the accelerator description using SystemC and SECDA.  
The compute module should be capable of performing L operations in parallel.
}

\end{document}